\documentclass[conference]{IEEEtran}
\IEEEoverridecommandlockouts
\usepackage{cite}
\usepackage{amsmath,amssymb,amsfonts}
\usepackage{algorithmic}
\usepackage{graphicx}
\usepackage{textcomp}
\usepackage{xcolor}
\def\BibTeX{{\rm B\kern-.05em{\sc i\kern-.025em b}\kern-.08em
    T\kern-.1667em\lower.7ex\hbox{E}\kern-.125emX}}

\usepackage{booktabs}
\usepackage{tabularx}
\usepackage{enumitem}

\usepackage{pifont}
\usepackage{fontawesome} % for fontawesome icons

\usepackage{colortbl}
\usepackage{array}
\usepackage{graphicx}
\usepackage{xcolor}

\usepackage{tcolorbox}

\newtcolorbox{vkquote}{
    colback=gray!10,
    colframe=gray!50,
    leftrule=2mm,
    arc=0pt,
    outer arc=0pt
}

\usepackage{url}
\usepackage{hyperref}

\usepackage[hyphenbreaks]{breakurl}

\usepackage{fancyhdr}
\newcommand{\quoteid}[1]{{\footnotesize #1}}

\newcommand{\change}[1]{\textcolor{black}{#1}}
\newcommand{\rebuttal}[1]{\textcolor{black}{#1}}
\newcommand{\new}[1]{\textcolor{black}{#1}}

\usepackage{lastpage}

\begin{document}

\title{Exploring Retrospective Meeting Practices \\and the Use of Data in Agile Teams}

\author{\IEEEauthorblockN{
\new{Alessandra Maciel Paz Milani\IEEEauthorrefmark{1},
Margaret-Anne Storey\IEEEauthorrefmark{1}, 
Vivek Katial\IEEEauthorrefmark{2} and
Lauren Peate\IEEEauthorrefmark{2}
}}
\IEEEauthorblockA{\IEEEauthorrefmark{1}
University of Victoria, Victoria, Canada}
\IEEEauthorblockA{\IEEEauthorrefmark{2}
\new{Multitudes, Auckland, New Zealand}}}

\maketitle

\begin{abstract}
Retrospectives are vital for software development teams to continuously enhance their processes and teamwork. Despite the increasing availability of objective data generated throughout the project and software development processes, many teams do not fully utilize this information in retrospective meetings. Instead, they often rely on subjective data, anecdotal insights and their memory. While some literature underscores the value of data-driven retrospectives, little attention has been given to the role data can play and the challenges of effectively incorporating objective project data into these meetings. To address this gap, we conducted a survey with 19 practitioners on retrospective meeting practices and how their teams gather and use subjective and objective data in their retrospectives. Our findings confirm that although teams routinely collect project data, they seldom employ it systematically during retrospectives. Furthermore, this study provides insights into retrospective practices by exploring barriers to project data utilization, including psychological safety concerns and the disconnect between data collection and meaningful integration of data into retrospective meetings. We close by considering preliminary insights that may help to mitigate these concerns and how future research might build on our paper findings to support the integration of project data into retrospective meetings, fostering a balance between human-centric reflections and data-driven insights.
\end{abstract}


%==================================================================
%==================================================================
\section{Introduction}

Retrospective meetings (retros) are structured gatherings held at the end of a project iteration where team members reflect on their recent work, identify successes, challenges, and areas for improvement, and plan actionable changes \cite{derby_agile_2006}. These meetings are a fundamental part of Agile methodology, which emphasizes iterative work cycles, continuous feedback, and flexibility to adapt to changing requirements \cite{highsmith_agile_2001, noauthor_manifesto_nodate}. 
The importance of retros lies in their ability to create an open environment for team members to voice their perspectives, encouraging collective problem-solving and transparency. This process not only improves the team’s workflow and productivity but also enhances communication, trust, and morale among team members \cite{manager_21_2021}. 

In a data-driven era, software development teams generate telemetry data from their tools, including commit patterns and test coverage from software repositories, collaboration metrics from communication platforms, and team satisfaction and experience data from regular surveys~\cite{singer_people_2017}.
Despite the increasing availability of this data \textit{and} the fact that most of the retro time is spent on gathering data \cite{dash_retrospectives_2019}, our internal studies with industry collaborators suggest that few teams consistently leverage a data-driven approach during retros. Instead, many rely on \rebuttal{subjective data---}anecdotal feedback or instinctive judgment---missing an opportunity to use the available \rebuttal{objective} data more effectively.

Incorporating \rebuttal{objective project} data into retros may offer benefits such as reducing memory biases by providing a \rebuttal{factual} record of events, enabling teams to reflect on their progress~\cite{bjarnason_evidence-based_2012}. Data also allows teams to challenge assumptions and explore issues from new perspectives, enriching retrospective meeting discussions.
Supporting these points, various studies have underscored the importance of using \rebuttal{project} data in Agile retros to enhance process improvement efforts \cite{erdogan_more_2018, matthies_experience_2021,bein_student_2023}. Still, there is limited discussion on the specific challenges and barriers associated with data utilization in retros. Hence, there is a need for further exploration into how teams can better integrate data into their reflection processes to drive measurable improvements~\cite{sharma_journey_2019}.

\rebuttal{This study addresses this disconnection between conducting retros and effectively using \rebuttal{project} data to inform decision-making and enhance processes. We designed a survey to explore retrospective practices and how teams gather and use data during retro meetings. Conducted in collaboration with an industry partner and its clients, the survey results (based on 19 teams' responses) contribute to a deeper understanding of the retro practices and the challenges surrounding using objective project data in Agile team retros. Inspired by that, we also offer preliminary recommendations to support teams' transition to a more data-informed retro meeting.}

This paper is structured as follows: Sec.~\ref{sec_background} provides background on retros\rebuttal{, types of project data, and data-driven approaches;} Sec.~\ref{sec_design} outlines our study design; Sec.~\ref{sec_findings} presents key findings; Sec.~\ref{sec_discussion} discusses \rebuttal{preliminary insights, limitations and future work}; Sec.~\ref{sec_conclusions} offers concluding remarks.

%==================================================================
%==================================================================
\section{Background and Related Work}
\label{sec_background}

In this section, we present a background on retro meetings, we clarify what project data entails, and we conclude by discussing related work on the use of data during retros.

\subsection{Retrospective Meetings: Structure, Tools, and Activities}
\label{sec_background_retro}

In Agile practices used in industry, such as Scrum~\cite{scrum, scrum_alliance},  retros play a crucial role in fostering a culture of ongoing improvement and collaboration by allowing teams to openly discuss both what went well and what needs adjustment, usually conducted at the end of every sprint (i.e., a fixed timeframe during which a Scrum team focuses on completing a defined amount of work).

Retros are typically structured around a five-step agenda that guides the team through reflecting on their recent work and planning improvements~\cite{derby_agile_2006}: (1) \textit{Setting the Stage}, which involves establishing the purpose and creating an open environment for discussions; (2) \textit{Gathering Data}, where the team collects insights and facts from the recent sprint iteration to build a shared understanding; (3) \textit{Generating Insights}, where the collected data is analyzed to identify patterns, issues, and opportunities; (4) \textit{Deciding What to Do}, during which the team formulates actionable plans and improvement strategies; and (5) \textit{Closing}, which aims to summarize the outcomes and reinforce commitments. 

To support the retro process, various tools and activities are used to facilitate productive discussions. Common digital tools such as Miro~\cite{miro} and Agile platforms such as Jira~\cite{jira} provide visual boards, templates, and analytics that help organize and review team insights. Additionally, interactive activities, as suggested in Retromat~\cite{retromat}, are frequently employed to encourage active participation. Ultimately, a successful retro relies heavily on team members’ active expression, underpinned by honesty, psychological safety\footnote{Previous studies outside the retro scope, also indicate psychological safety in the workplace as the most critical factor for team success~\cite{mccausland_creating_2023,nyt_online}.}, openness, and a commitment to improvement~\cite{manager_21_2021}.

\subsection{\rebuttal{Project Data and Frameworks}}
\label{sec_background_data}

Broadly, data can be categorized into ``hard'' data and ``soft'' data \cite{derby_agile_2006,matthies_experience_2021}. Hard data includes quantitative metrics \rebuttal{about the project} such as code commits, test results, bug counts, review data, and deployment frequencies—tangible indicators that provide an objective view of the development process. On the other hand, soft data encompasses qualitative inputs like team sentiment, feedback from developers, and collaboration experiences, which offer a subjective but valuable perspective on team dynamics.

Among trending frameworks for using these types of data are SPACE~\cite{forsgren_space_2021}, that evaluates team productivity and satisfaction across multiple dimensions, and DORA metrics~\cite{google_2024_dora} that focuses on four key metrics to gauge software delivery performance (\textit{change lead time}, \textit{deployment frequency}, \textit{mean time to restore}, and \textit{change failure rate}). 
Despite their growing popularity and the recognized importance of data-driven approaches in Software Engineering~\cite{yalciner_data-driven_2024} and Agile methodologies~\cite{fawzy_exploring_2024, almeida_perceived_2023, biesialska_big_2021, ram_data-driven_2022, matthies_towards_2019}, effectively integrating these frameworks and metrics into retros remains an opportunity for many Agile teams, suggesting the need for deeper exploration into the problem understanding of data-driven practices in retros. 
\rebuttal{In the following subsection, we provide references that explore the use of hard data and data-driven approaches in retrospectives.}

\subsection{\rebuttal{Use of Project Data during Retrospective Meetings}}
\label{sec_background_rw}

Erdoğan \textit{et al.}~\cite{erdogan_more_2018} explored how statistical analysis can improve the effectiveness of sprint retros. They noted that ``gathering correct data requires trust, respect, collaboration, and motivation within the team,'' emphasizing the importance of team dynamics in adopting data-driven approaches. Although statistical analysis has a positive impact on process improvement and team velocity, challenges to adoption remain. These include defining which analyses are most relevant for retro insights and determining practical methods for implementing these analyses in team settings.

Later, other studies from Matthies \textit{et al.}~\cite{matthies_feedback_2019, matthies_mining_2020, matthies_playing_2020} explored incorporating insights from mining software repositories (e.g., GitHub) into retro agendas. 
In a 2021 study \cite{matthies_experience_2021}, they reviewed Retromat~\cite{retromat} activities, specifically examining the \textit{Gathering Data} phase~\cite{derby_agile_2006} and the use of project data. Their analysis found that 86\% of Retromat's data-gathering activities lack explicit connections to project data, relying mainly on participant perceptions. 
To address this, they proposed modifications to five additional activities to better integrate project data, supporting more evidence-based decision-making in retros.

Although Matthies \textit{et al.} (above) were among the first to discuss using project data in retros, they did not address the challenges of adoption, which is done by Bein~\cite{bein_student_2023}, in a related study to ours. The author conducted interviews with seven agile practitioners and identified a lack of guidelines and theories for integrating data analysis in retros, both in industry and academic research.
Most of their interviewees recognize the potential benefits of using project data in retros. However, they also highlighted challenges, such as deciding which project artifacts to use, formulating relevant questions, and implementing analyses (similarly to what \cite{erdogan_more_2018} pointed out). Their concerns also included the high effort required and fears that data might be misused for surveillance or assigning blame.

The research summarized above provides preliminary insights on some of the benefits and challenges of using data in retros. We need to confirm and deepen these insights, and learn how to address the barriers and challenges of using \rebuttal{objective project data} in Agile team retros.

%==================================================================
%==================================================================
\section{Study Design}
\label{sec_design}

\rebuttal{To answer our research question: \textit{What practices do Agile teams follow when conducting retrospective meetings, and how do they gather and utilize data beforehand and during these meetings?},} we designed a survey in collaboration with an industry partner\new{---Multitudes\footnote{\url{https://www.multitudes.com}}}, a software company whose main product is an engineering insights platform for team efficiency and well-being. 
The survey was selected as the data collection method considering the fact respondents can take the online survey at their convenience, reducing the need for scheduling and allowing for flexibility in different time zones. In the next subsections, we describe the data collection, the recruitment of respondents, \rebuttal{and the data analysis process.}

\subsection{Data Collection}

The survey was designed to be concise (with a completion time of under 15 minutes). Respondents were first presented with an informed consent form and study information,\footnote{The study protocol was approved by the Human Research Ethics Office at the University \new{of Victoria, application number: 23-0604}.} followed by a total of 14 questions. 
The questions inquired about:
\rebuttal{(a) Team Profiles (size and experience);
(b) Project Sprint length;
(c) Retro meeting practices (frequency, duration, configuration and tooling);
(d) Project data tracking and usage in preparation for \textit{and} during retro meetings.}
\rebuttal{In this paper, the survey questions are referenced as Q2 to Q13. A table containing detailed descriptions of all questions, along with team responses to the close-ended questions, was included in the supplementary material\footnote{\label{supmat}See supplementary material \url{https://doi.org/10.5281/zenodo.14619024}}.}

\subsection{Recruitment and Respondents}

Nineteen responses were received for the survey during March and April 2024. The industry partner distributed the invitation via email and the Slack channels, targeting some of their clients. 
The unit of analysis of this study is the team. Thus, each response corresponds to a different team, indicated during the presentation of results as T01 to T19\textsuperscript{\ref{supmat}}.
Approximately 100 teams were reached through these communications, resulting in a participation rate of 19\%.

\subsection{Data Analysis}

The data analysis process involved reviewing responses from the survey, which were a mix of closed and open ended questions. 
First, one researcher summarized the quantitative data from the responses related to the team demographics and practices, such as the percentage of teams conducting retro meetings every two weeks. 
Next, this researcher categorized the responses of the open questions and assigned codes (e.g., ``retro data'' or ``retro tool''---for a total of 21 inductive codes). 

Later, the same researcher summarized common trends and main insights derived from the coding process in key themes that were discussed with the entire research project team. 
During these discussions, some initial themes were merged and their titles revisited, which resulted in 12 final themes (\rebuttal{eight themes described in Sec.~\ref{sub:thematic1} and four in Sec.~\ref{sub:thematic2}}).

%==================================================================
%==================================================================
\section{Findings}
\label{sec_findings}

First, we present the overall \textit{Team Demographics and Their Practices} (based on questions Q2--Q10\textsuperscript{\ref{supmat}}). Next, we continue the presentation of findings from the thematic analysis of the final open questions (Q11--Q13\textsuperscript{\ref{supmat}}), starting with the insights on the \textit{Current Retros Practice}, and concluding with the \textit{Challenges and Barriers to Using Data in Retros}.

\subsection{Team Demographics and Their Practices}

Our analysis of the team demographics and their retro practices revealed that most teams who answered our survey conduct retros every two weeks (16, 84\%), with meetings generally lasting 60 minutes (12, 63\%). 
The teams vary in size, ranging from five to 16 members, and most teams conduct their retros remotely or online (16, 84\%), but some use hybrid formats (3, 16\%). Team composition changes were reported to occur between two weeks to nine months prior to the survey. Notably, the teams (14, 74\%) have at least one year of experience with retros, with some reporting five to eight years of experience. There is also variation in sprint duration, with many teams operating on a two-week cycle (13, 68\%), while others use different sprint duration or no sprints at all (e.g., Kanban~\cite{kanban_online} or \textit{ad hoc}). 

The survey respondents indicated the use of the following tools during or in preparation to their retro meetings: 
\begin{enumerate}[label=\alph*.]
    \item Engineering team metrics \& analytics: GetDx (DX Cloud); GitHub; Google Sheets and Spreadsheets (data from other tools); in-house dashboards; Jira; Linear.app; LinearB; Multitudes; Nave; Notion/in-house guide.
    \item Task management or issue tracking data: Asana; ClickUp; Google Sheets; Jira; Linear.app; Notebook; Youtrack.
    \item Whiteboards (Physical or Online): ClickUp; FigJam; Miro; Mutal; RemoteRetro; Whiteboard; Whimsical.
    \item Others (\textit{open to add any tools unmentioned for the previous categories a--c}): EasyRetro; Google Looker; Notion; parabol.co; Retrium; Slack; Zoom.
\end{enumerate}
It is noteworthy that some answered ``no tools are used'' during their retros, as well as, other tools not mentioned on this list are used during their data gathering or other team activities (mentioned by them for other questions).

As a final remark for this section, most respondents see retros as highly valuable for improving team collaboration. 
In regard to \textit{how the team works together} (Q6-A), the majority (9, 47\%) consider retros to be \textit{VERY important}, while six (32\%) find them \textit{Important}. 
However, when considering \textit{the work the team does} (Q6-B), opinions are more varied. While seven (37\%) consider retros \textit{Important}, only two (11\%) view them as \textit{VERY important}. A larger portion of respondents (5, 26\%) rated retros as \textit{Moderately important}, and four (21\%) said they are only \textit{Slightly important}. This suggests that while retros are seen as valuable for teamwork, their perceived impact on the actual work being done is somewhat less significant.

\subsection{Current Retrospective Practice}
\label{sub:thematic1}

In the following subsections, we present eight themes for the current retro practice that emerged from our thematic analysis. In support of each theme, we share at least one representative or noteworthy quote from the respondents.

\subsubsection{\textbf{The Retro Starts Before the Meeting Start}}
\label{theme1}
Preparation involves setting up tools, gathering metrics (if used), and allowing team members to add topics in advance, which helps capture thoughts or concerns for the retro.
\begin{itemize}[leftmargin=0.4cm]
    \item [\scriptsize \faComment] \textit{As a remote async team, we allow people to add discussion points to the retro anytime leading up to the retro.} \quoteid{[Q12-T05]}
    \item [\scriptsize \faComment] \textit{Before starting, I (EM) usually get a Retro board template \rebuttal{(...)} and make sure everything is ready (...). I also prepare the (...) metrics to ensure the Retro runs smoothly.} \quoteid{[Q12-T17]}    
\end{itemize}

\subsubsection{\textbf{Custom Retro Board and Different Tools}} Most teams \change{(15, 79\%)} often use custom retro boards and tools to facilitate structured conversations and capture discussion points.  
\begin{itemize}[leftmargin=0.4cm]
    \item [\scriptsize \faComment] \textit{It is Free-form, whiteboard in Whimsical is provided early for capturing thoughts.} \quoteid{[Q12-T03]}
    \item [\scriptsize \faComment] \textit{Retros are usually structured around a traditional what worked well / didn't work well / do differently board in FigJam. We try to mix up the style with different templates FigJam, Miro, or resources on the internet.} \quoteid{[Q12-T14]} 
\end{itemize}

\subsubsection{\textbf{Creative and Informal Approaches}}
Some teams have introduced more casual or creative elements into their retros, such as using fun metaphors or non-traditional structures to encourage participation and reflection. This highlights a desire for flexibility and innovation in how retros are run. 
\begin{itemize}[leftmargin=0.4cm]
    \item [\scriptsize \faComment] \textit{Our team meets weekly to just talk openly... this is a very open session and has very little structure.} \quoteid{[Q13-T06]} 
    \item [\scriptsize \faComment] \textit{We started sharing `what animal would you use to describe how you felt during the sprint' \rebuttal{(...)}} \quoteid{[Q13-T15]} 
    \item [\scriptsize \faComment] \textit{Our retro is actively being changed \rebuttal{(...)}} \quoteid{[Q12-T18]} 
\end{itemize}

\subsubsection{\textbf{Retro Used to Support Just-in-Time Learning}}
Two teams explicitly mentioned the aim to foster a growth mindset by encouraging failure as an opportunity for learning and only holding retros when there is a clear need or benefit, demonstrating a strategic and intentional approach to retros. 
\begin{itemize}[leftmargin=0.4cm]
    \item [\scriptsize \faComment] \textit{We try to have a growth mindset and support/encourage failure and the learning attached to failing. We really only have retrospectives when there is a real need or something to be learned/improved.} \quoteid{[Q13-T06]} 
\end{itemize}    

\subsubsection{\textbf{Retro is Important but still a Work in Progress}} 
Several respondents indicated that while retros are important, their current execution is lacking, leading to a sense that retros could be more effective or are even a ``waste of time'' due to poor implementation. 
This helps to explain the differing perceptions of the importance of retros in terms of teamwork and the work done by the team (related to Q6 introduced earlier). 
\begin{itemize}[leftmargin=0.4cm]
    \item [\scriptsize \faComment] \textit{We follow Shape Up\footnote{\url{https://agilefirst.io/what-is-shape-up/}} and deliberately choose to bring in Retros from agile methods as one of the things we liked from Agile/Scrum.} \quoteid{[Q12-T05]}
    \item [\scriptsize \faComment] \textit{Retros are important; however, \rebuttal{(...)} the team's execution of retros right now is poor and \rebuttal{(...)} a waste of time.}\quoteid{[Q13-T18]}    
\end{itemize} 

\subsubsection{\textbf{Team Structure or Team Size Effect on Retros}}
Teams are experimenting with restructuring and breaking into smaller groups for more focused meetings. This theme reflects an effort to improve teamwork and decision-making through organizational changes (one team even noted this action of having smaller teams came out of their previous retro reflection).%(subject emerged from their past retros). 
\begin{itemize}[leftmargin=0.4cm] 
    \item [\scriptsize \faComment] \textit{Our retros split into smaller groups of 3-5 at random. The small retro team is responsible for implementing the actions---they [don't] need any extra alignment, permission or an ok from leadership.} \quoteid{[Q12-T05]}
    \item [\scriptsize \faComment]\textit{We have been doing retros as a larger engineering team for years but recently reorganized into two smaller teams.} \quoteid{[Q13-T16]}    
\end{itemize} 

\subsubsection{\textbf{\rebuttal{Use of Data in Retros}}}
\label{theme7}
\rebuttal{The project data tracked by teams varies widely, encompassing operational metrics such as velocity and story points, as well as collaboration metrics like pull request feedback and code review gaps. This objective data is often drawn from multiple tools and reviewed at varying frequencies, depending on the metric's nature.
Despite some teams incorporating specific engineering metrics into retrospectives, the use of objective data in these meetings remains limited. Most teams prioritize open discussions and personal reflections (these different approaches are further explored in the next and final theme).} 
\begin{itemize}[leftmargin=0.4cm]
    \item [\scriptsize \faComment] \textit{We keep track of a lot of data and use it to a varying degree. Typically, data is not used in a retro for this team.} \quoteid{[Q11-T19]} 
    \item [\scriptsize \faComment] \rebuttal{\textit{Engineering Metrics (...) Execution Metrics (...). During the meeting, we go over the Retro board (around 30 min for that), and in the last 15 minutes, we present and discuss the team's metrics and opportunities for improvement.} \quoteid{[Q12-P17]}}    
\end{itemize} 

\subsubsection{\textbf{Human-Centric \rebuttal{and} Data-Driven Approaches}} 
A common theme is the emphasis on the human aspect of retros, with teams focusing on empathy, well-being, and workplace satisfaction over purely data-driven processes. 
\rebuttal{While only six explicitly reported they bring objective project data for retro discussion, nine explicitly mentioned the use of subjective data. Thus, f}or many, retros are more about open communication (i.e., more open-ended, personal ``how did you feel'' style) than reviewing hard metrics (i.e., engineering metrics or execution metrics). 
\begin{itemize}[leftmargin=0.4cm]
    \item [\scriptsize \faComment] \textit{Our retrospectives are mostly based on what people felt went right or wrong, of things we think we may need to improve, celebrate, change, or do differently. It isn't very metric-centric.} \quoteid{[Q12-T08]}
    \item [\scriptsize \faComment] \textit{\rebuttal{[Retros]} have more of an [empathetic] approach than a data-led approach. (...) We aim for the retro to be a safe space for people to share thoughts and ideas.} \quoteid{[Q13-T10]}    
\end{itemize} 

\subsection{Challenges and Barriers to Using Data in Retrospective}
\label{sub:thematic2} 

Some survey responses highlight challenges such as under-use of available data, a gap between management and team-level data usage, and a lack of integration of valuable metrics into routine processes.
To provide further examples of these issues, we introduce four themes that emerged from our data analysis in the following subsections. 

\subsubsection{\textbf{Preference for Simple Tracking Methods and Few Metrics}} \label{theme9_challenge1}
Although detailed metrics are available through their tools and platforms, some teams prefer to rely on simpler tracking methods, such as boards or spreadsheets, indicating a reluctance or challenge in adopting more advanced data-driven approaches during retros. 
\begin{itemize}[leftmargin=0.4cm]
    %\scriptsize
    \item [\scriptsize \faComment] \textit{We mostly just look at our board and try to ensure things are moving and that we're minimizing and managing any rollovers well.} \quoteid{[Q11-T16]}
    \item [\scriptsize \faComment] \textit{Our primary focus is overall sprint delivery; we track planned points, and delivered points (...). We keep track of this in a very simple spreadsheet.} \quoteid{[Q11-T10]}
\end{itemize} 

\subsubsection{\textbf{Data Tracking Without Integration into Processes}}
\label{theme10_challenge2}
Teams track a variety of metrics, but there seems to be a lack of integration of this objective data into regular processes like sprint planning or retros, limiting its practical use.
\begin{itemize}[leftmargin=0.4cm]
    \item [\scriptsize \faComment] \textit{There is other data that we keep track of but that we have not integrated into our processes directly.} \quoteid{[Q11-T16]}       
    \item [\scriptsize \faComment] \textit{We've been doing retros for years but really haven't done a great job of integrating metrics into our processes.}~\quoteid{[Q13-T16]}     
\end{itemize} 

\subsubsection{\textbf{Asymmetric Data Usage and Access by Different Roles}}
\label{theme11_challenge3}
Data access and usage are role-specific and hierarchical. In some cases, managers or team leads track and use detailed data for decision-making,  
while the broader team does not engage with this data during meetings or collaborative sessions (relying only on higher-level information).
\begin{itemize}[leftmargin=0.4cm]
    \item [\scriptsize \faComment] \textit{I use data myself as a manager and team leader to monitor trends and try to address areas of improvement but we do not look at data directly as a team in ceremonies.} \quoteid{[Q11-T16]}
    \item [\scriptsize \faComment] \textit{Most of the team do check their own personal sprint delivery before retro \rebuttal{(...)}, but it's not mandatory. As a team lead I don't generally bring team-level statistics into the retro meeting unless there's a critical problem.} \quoteid{[Q12-T10]}   
\end{itemize} 

\subsubsection{\textbf{Psychological Safety}}
\label{theme12_challenge4}
While metrics are acknowledged as valuable, some responses indicate a sense of caution around metrics, leading to a defensive mindset, where team members feel they need to justify their performance rather than focus on growth. This suggests the importance of \rebuttal{creating} a safe collaborative place when introducing data-driven approaches.
\begin{itemize}[leftmargin=0.4cm]
    \item [\scriptsize \faComment] \textit{We rely on being small enough and high-trust enough to just say 'hey, this review is taking a while' rather than going over metrics.} \quoteid{[Q13-T08]}
    \item [\scriptsize \faComment] \textit{This can sometimes lead to what feels like a need for the team to `defend' or `explain' behaviour that forms trends/results.} \quoteid{[Q13-T07]}
\end{itemize} 

This theme concludes our key findings.
Below, we discuss the study's implications \rebuttal{key insights}.

%==================================================================
%==================================================================
\section{\rebuttal{Discussion}}
\label{sec_discussion}

\rebuttal{Building on our survey findings (Sec.~\ref{sec_findings}) alongside other observations and prior studies (Sec.~\ref{sec_background_rw}), we offer additional perspectives on the use of project data in retros and outline preliminary considerations in the next subsection. We then discuss the limitations of our study and future work.}

\subsection{\rebuttal{Actionable Insights}}

\rebuttal{Our findings confirm the limited use of project data in retros (Sec.~\ref{theme7}), as noted in prior studies.} 
However, unlike previous work, our respondents did not report challenges with selecting or analyzing data (as in \cite{erdogan_more_2018, bein_student_2023}), nor the need for additional supporting activities (as in \cite{matthies_experience_2021}). 
Although we did not explicitly ask about these challenges in our survey, this may be because their available tool (developed by our industry partner) provides data summaries and team insights. 

Considering the challenges and barriers reported \rebuttal{in our study}, the reliance on simple metrics (\rebuttal{Sec.~\ref{theme9_challenge1}}) may stem from time constraints and ease of access, as more complex metrics may require more time to gather and interpret. This leads to \rebuttal{our first actionable insight:}

\begin{itemize}[leftmargin=0.4cm]
    \item [\scriptsize \faBookmark] \textit{Make it easier for practitioners \rebuttal{to obtain a range of insights from their subjective and objective data (i.e., not only soft data about team sentiment or simple hard data such as code commits, but both are needed)}.}
\end{itemize} 

While the issue of data tracking without integration (\rebuttal{Sec.~\ref{theme10_challenge2}}) could result from the absence of regular reminders, \rebuttal{this may lead} teams to overlook valuable insights on dashboards. \rebuttal{Therefore:}

\begin{itemize}[leftmargin=0.4cm]
    \item [\scriptsize \faBookmark] \textit{Proactively remind practitioners about the key insights---do not wait for them to dive into a dashboard.}
\end{itemize} 
 
In our internal studies before the survey reported in this study, practitioners expressed a preference for a ``tool for better discussions'' \rebuttal{instead of simply} a ``new retro tool.''
\rebuttal{Given that retros begin before the actual meeting (Sec.~\ref{theme1}), alternative approaches for team communication should be considered.}

\begin{itemize}[leftmargin=0.4cm]
    \item [\scriptsize \faBookmark] \textit{Make it easier for practitioners to discuss \rebuttal{their subjective and objective project data} in advance of the session.}
\end{itemize}

Lastly, the importance of psychological safety (\rebuttal{Sec.~\ref{theme12_challenge4}}) likely underlies several other findings. Leaders may be more inclined to \rebuttal{use objective project} data in retros (\rebuttal{Sec.~\ref{theme11_challenge3}}) through transparent sharing if they felt confident that their teams would engage openly and comfortably with the data.

\begin{itemize}[leftmargin=0.4cm]
    \item [\scriptsize \faBookmark] \textit{\rebuttal{All data} should be used with trust and transparency in mind.}
\end{itemize}

Even if it is high-level or preliminary, we chose these four actionable insights to provide an initial direction for practitioners. Our insights can also inspire further studies, such as directly asking participants about their concerns in using project data during retros and which opportunities could be explored to support them. Additional considerations on this study's limitation are presented in the following subsection, and thoughts on further investigations in the final subsection.

\subsection{Study Limitations}

A key limitation of our study is that the sample size means our findings may only represent some Agile teams. Further, our respondents were from  organizations that collaborate and use the tool provided by our industry partner, which may introduce a bias.
Another constraint is the survey's limited questions on organizational culture and Agile maturity levels, which could also influence a data-driven adoption in retros. Still, while no questions explicitly addressed the pros and cons of project data use, the challenges and barriers mentioned in open-ended responses were even more insightful.

\subsection{\rebuttal{Future Work}}

Future research and tool development should continue to seek to understand the retros practices, the role of data and how to foster a balanced approach that combines empathetic feedback and subjective data with objective project data, enhancing retros as both reflective and data-informed meetings. 
For example, a data analysis tool could include an annotation feature for asynchronous team discussions, allowing notes on data points (chart in the teams' dashboard) to later be added to the retro agenda. 
This is one potential intervention for future studies, among many others to explore. We hope our findings and preliminary insights inspire continued investigations.

%==================================================================
%==================================================================
\section{Conclusion}
\label{sec_conclusions}

This study explored retro practices and challenges associated with incorporating project data into retros by Agile teams. Key challenges include psychological safety concerns and poor integration of metrics. Addressing these challenges requires tools and strategies fostering a collaborative environment where data is used to empower teams rather than intimidate them. Future research should explore integrating objective project data during retro meetings to support balancing human-centric reflections with data-driven insights.

%------------------------------------------------------------------------------
%------------------------------------------------------------------------------
% Acknowledgment
%------------------------------------------------------------------------------
%------------------------------------------------------------------------------
\section*{Acknowledgment}
\new{The authors would like to thank and report the financial support provided for project research during this study by the Natural Sciences and Engineering Research Council of Canada (NSERC).
The authors also thank all the participants of this study and the reviewers of this manuscript.}

%------------------------------------------------------------------------------
%------------------------------------------------------------------------------
% References
%------------------------------------------------------------------------------
%------------------------------------------------------------------------------
% \section*{References}
\bibliographystyle{IEEEtran}
\bibliography{references.bib}

% Generated by IEEEtran.bst, version: 1.14 (2015/08/26)
\begin{thebibliography}{10}
\providecommand{\url}[1]{#1}
\csname url@samestyle\endcsname
\providecommand{\newblock}{\relax}
\providecommand{\bibinfo}[2]{#2}
\providecommand{\BIBentrySTDinterwordspacing}{\spaceskip=0pt\relax}
\providecommand{\BIBentryALTinterwordstretchfactor}{4}
\providecommand{\BIBentryALTinterwordspacing}{\spaceskip=\fontdimen2\font plus
\BIBentryALTinterwordstretchfactor\fontdimen3\font minus \fontdimen4\font\relax}
\providecommand{\BIBforeignlanguage}[2]{{%
\expandafter\ifx\csname l@#1\endcsname\relax
\typeout{** WARNING: IEEEtran.bst: No hyphenation pattern has been}%
\typeout{** loaded for the language `#1'. Using the pattern for}%
\typeout{** the default language instead.}%
\else
\language=\csname l@#1\endcsname
\fi
#2}}
\providecommand{\BIBdecl}{\relax}
\BIBdecl

\bibitem{derby_agile_2006}
E.~Derby and D.~Larsen, \emph{Agile retrospectives: making good teams great}.\hskip 1em plus 0.5em minus 0.4em\relax Raleigh, NC: Pragmatic Bookshelf, 2006, oCLC: ocm71756468.

\bibitem{highsmith_agile_2001}
\BIBentryALTinterwordspacing
J.~Highsmith and A.~Cockburn, ``Agile software development: the business of innovation,'' \emph{Computer}, vol.~34, no.~9, pp. 120--127, Sep. 2001. [Online]. Available: \url{https://ieeexplore.ieee.org/document/947100}
\BIBentrySTDinterwordspacing

\bibitem{noauthor_manifesto_nodate}
\BIBentryALTinterwordspacing
``Manifesto for {Agile} {Software} {Development}.'' [Online]. Available: \url{https://agilemanifesto.org/}
\BIBentrySTDinterwordspacing

\bibitem{manager_21_2021}
\BIBentryALTinterwordspacing
G.~Low, ``\BIBforeignlanguage{en-US}{21 {Sprint} {Retrospective} {Ideas} {To} {Get} {Quiet} {Project} {Teams} {Talking}},'' Aug. 2021. [Online]. Available: \url{https://thedigitalprojectmanager.com/projects/leadership-team-management/get-quiet-teams-talking-agile-retrospectives/}
\BIBentrySTDinterwordspacing

\bibitem{singer_people_2017}
L.~Singer, M.-A. Storey, F.~Figueira~Filho, A.~Zagalsky, and D.~M. German, ``\BIBforeignlanguage{en}{People {Analytics} in {Software} {Development}},'' in \emph{\BIBforeignlanguage{en}{Grand {Timely} {Topics} in {Software} {Engineering}}}, J.~Cunha, J.~P. Fernandes, R.~LXmmel, J.~Saraiva, and V.~Zaytsev, Eds.\hskip 1em plus 0.5em minus 0.4em\relax Cham: Springer International Publishing, 2017, pp. 124--153.

\bibitem{dash_retrospectives_2019}
\BIBentryALTinterwordspacing
S.~N. Dash, ``\BIBforeignlanguage{en-US}{Retrospectives and {Intraspectives} for {Agile} {Practitioners}},'' Oct. 2019. [Online]. Available: \url{https://mpug.com/retrospectives-and-intraspectives-for-agile-practitioners/}
\BIBentrySTDinterwordspacing

\bibitem{bjarnason_evidence-based_2012}
E.~Bjarnason and B.~Regnell, ``\BIBforeignlanguage{en}{Evidence-{Based} {Timelines} for {Agile} {Project} {Retrospectives} - {A} {Method} {Proposal}},'' in \emph{\BIBforeignlanguage{en}{Agile {Processes} in {Software} {Engineering} and {Extreme} {Programming}}}, C.~Wohlin, Ed.\hskip 1em plus 0.5em minus 0.4em\relax Berlin, Heidelberg: Springer, 2012, pp. 177--184.

\bibitem{erdogan_more_2018}
\BIBentryALTinterwordspacing
O.~Erdoğan, M.~E. Pekkaya, and H.~Gök, ``\BIBforeignlanguage{en}{More effective sprint retrospective with statistical analysis},'' \emph{\BIBforeignlanguage{en}{Journal of Software: Evolution and Process}}, vol.~30, no.~5, p. e1933, May 2018. [Online]. Available: \url{https://onlinelibrary.wiley.com/doi/10.1002/smr.1933}
\BIBentrySTDinterwordspacing

\bibitem{matthies_experience_2021}
C.~Matthies and F.~Dobrigkeit, ``Experience vs data: A case for more data-informed retrospective activities,'' in \emph{Lean and Agile Software Development}, A.~Przyby{\l}ek, J.~Miler, A.~Poth, and A.~Riel, Eds.\hskip 1em plus 0.5em minus 0.4em\relax Cham: Springer International Publishing, 2021, pp. 130--144.

\bibitem{bein_student_2023}
\BIBentryALTinterwordspacing
L.~Bein, ``Student {Research} {Abstract}: {Why} and {Where} {Software} {Developers} are (not) using {Project} {Data} in {Agile} {Retrospectives},'' in \emph{Proceedings of the 38th {ACM}/{SIGAPP} {Symposium} on {Applied} {Computing}}, ser. {SAC} '23.\hskip 1em plus 0.5em minus 0.4em\relax New York, NY, USA: Association for Computing Machinery, Jun. 2023, pp. 1060--1063. [Online]. Available: \url{https://dl.acm.org/doi/10.1145/3555776.3577205}
\BIBentrySTDinterwordspacing

\bibitem{sharma_journey_2019}
\BIBentryALTinterwordspacing
V.~S. Sharma, R.~Mehra, S.~Podder, and A.~P. Burden, ``A {Journey} {Towards} {Providing} {Intelligence} and {Actionable} {Insights} to {Development} {Teams} in {Software} {Delivery},'' in \emph{2019 34th {IEEE}/{ACM} {International} {Conference} on {Automated} {Software} {Engineering} ({ASE})}, Nov. 2019, pp. 1214--1215, iSSN: 2643-1572. [Online]. Available: \url{https://ieeexplore.ieee.org/document/8952441}
\BIBentrySTDinterwordspacing

\bibitem{scrum}
\BIBentryALTinterwordspacing
``Home {\textbar} {Scrum} {Guides}.'' [Online]. Available: \url{https://scrumguides.org/}
\BIBentrySTDinterwordspacing

\bibitem{scrum_alliance}
``The {State} of {Scrum} {Report},'' \url{https://www.scrumalliance.org/ScrumRedesignDEVSite/media/ScrumAllianceMedia/Files%20and%20PDFs/State%20of%20Scrum/2017-SoSR-Final-Version-(Pages).pdf}.

\bibitem{miro}
\BIBentryALTinterwordspacing
``Miro.'' [Online]. Available: \url{https://miro.com/}
\BIBentrySTDinterwordspacing

\bibitem{jira}
\BIBentryALTinterwordspacing
``Jira.'' [Online]. Available: \url{https://www.atlassian.com/software/jira}
\BIBentrySTDinterwordspacing

\bibitem{retromat}
\BIBentryALTinterwordspacing
``\BIBforeignlanguage{en-US}{Retromat {Miroboard} {Mega} {Template} {Retrospectives}}.'' [Online]. Available: \url{https://retromat.org/blog/retromat-miroboard-mega-template/}
\BIBentrySTDinterwordspacing

\bibitem{mccausland_creating_2023}
\BIBentryALTinterwordspacing
T.~McCausland, ``\BIBforeignlanguage{en}{Creating {Psychological} {Safety} in the {Workplace}},'' \emph{\BIBforeignlanguage{en}{Research-Technology Management}}, vol.~66, no.~2, pp. 56--58, Mar. 2023. [Online]. Available: \url{https://www.tandfonline.com/doi/full/10.1080/08956308.2023.2164439}
\BIBentrySTDinterwordspacing

\bibitem{nyt_online}
\BIBentryALTinterwordspacing
C.~Duhigg, ``What {G}oogle learned from its quest to build the perfect team.'' [Online]. Available: \url{https://www.nytimes.com/2016/02/28/magazine/what-google-learned-from-its-quest-to-build-the-perfect-team.html}
\BIBentrySTDinterwordspacing

\bibitem{forsgren_space_2021}
\BIBentryALTinterwordspacing
N.~Forsgren, M.-A. Storey, C.~Maddila, T.~Zimmermann, B.~Houck, and J.~Butler, ``\BIBforeignlanguage{en}{The {SPACE} of {Developer} {Productivity}: {There}'s more to it than you think.}'' \emph{\BIBforeignlanguage{en}{Queue}}, vol.~19, no.~1, pp. 20--48, Feb. 2021. [Online]. Available: \url{https://dl.acm.org/doi/10.1145/3454122.3454124}
\BIBentrySTDinterwordspacing

\bibitem{google_2024_dora}
\BIBentryALTinterwordspacing
Google, ``\BIBforeignlanguage{en-US}{2024 {State} of {DevOps} {Report}}.'' [Online]. Available: \url{https://cloud.google.com/resources/devops/state-of-devops}
\BIBentrySTDinterwordspacing

\bibitem{yalciner_data-driven_2024}
A.~Yalçıner, A.~Dikici, and E.~Gökalp, ``\BIBforeignlanguage{en}{Data-{Driven} {Software} {Engineering}: {A} {Systematic} {Literature} {Review}},'' in \emph{\BIBforeignlanguage{en}{Systems, {Software} and {Services} {Process} {Improvement}}}, M.~Yilmaz, P.~Clarke, A.~Riel, R.~Messnarz, C.~Greiner, and T.~Peisl, Eds.\hskip 1em plus 0.5em minus 0.4em\relax Cham: Springer Nature Switzerland, 2024, pp. 19--32.

\bibitem{fawzy_exploring_2024}
\BIBentryALTinterwordspacing
A.~Fawzy, A.~Tahir, M.~Galster, and P.~Liang, ``Exploring {Data} {Management} {Challenges} and {Solutions} in {Agile} {Software} {Development}: {A} {Literature} {Review} and {Practitioner} {Survey},'' Jul. 2024, arXiv:2402.00462 [cs]. [Online]. Available: \url{http://arxiv.org/abs/2402.00462}
\BIBentrySTDinterwordspacing

\bibitem{almeida_perceived_2023}
\BIBentryALTinterwordspacing
F.~Almeida and P.~Carneiro, ``\BIBforeignlanguage{en}{Perceived {Importance} of {Metrics} for {Agile} {Scrum} {Environments}},'' \emph{\BIBforeignlanguage{en}{Information}}, vol.~14, no.~6, p. 327, Jun. 2023. [Online]. Available: \url{https://www.mdpi.com/2078-2489/14/6/327}
\BIBentrySTDinterwordspacing

\bibitem{biesialska_big_2021}
\BIBentryALTinterwordspacing
K.~Biesialska, X.~Franch, and V.~Muntés-Mulero, ``Big {Data} analytics in {Agile} software development: {A} systematic mapping study,'' \emph{Information and Software Technology}, vol. 132, p. 106448, Apr. 2021. [Online]. Available: \url{https://www.sciencedirect.com/science/article/pii/S0950584920301981}
\BIBentrySTDinterwordspacing

\bibitem{ram_data-driven_2022}
\BIBentryALTinterwordspacing
P.~Ram, ``\BIBforeignlanguage{eng}{Data-driven process improvement in agile software development : an industrial multiple-case study},'' Oct. 2022. [Online]. Available: \url{https://oulurepo.oulu.fi/handle/10024/36880}
\BIBentrySTDinterwordspacing

\bibitem{matthies_towards_2019}
\BIBentryALTinterwordspacing
C.~Matthies and G.~Hesse, ``Towards {Using} {Data} to {Inform} {Decisions} in {Agile} {Software} {Development}: {Views} of {Available} {Data},'' in \emph{Proceedings of the 14th {International} {Conference} on {Software} {Technologies}}, 2019, pp. 552--559, arXiv:1907.12959 [cs]. [Online]. Available: \url{http://arxiv.org/abs/1907.12959}
\BIBentrySTDinterwordspacing

\bibitem{matthies_feedback_2019}
\BIBentryALTinterwordspacing
C.~Matthies, ``Feedback in {Scrum}: {Data}-{Informed} {Retrospectives},'' in \emph{2019 {IEEE}/{ACM} 41st {International} {Conference} on {Software} {Engineering}: {Companion} {Proceedings} ({ICSE}-{Companion})}, May 2019, pp. 198--201, iSSN: 2574-1934. [Online]. Available: \url{https://ieeexplore.ieee.org/abstract/document/8802669}
\BIBentrySTDinterwordspacing

\bibitem{matthies_mining_2020}
\BIBentryALTinterwordspacing
C.~Matthies, F.~Dobrigkeit, and G.~Hesse, ``Mining for {Process} {Improvements}: {Analyzing} {Software} {Repositories} in {Agile} {Retrospectives},'' in \emph{Proceedings of the {IEEE}/{ACM} 42nd {International} {Conference} on {Software} {Engineering} {Workshops}}, ser. {ICSEW}'20.\hskip 1em plus 0.5em minus 0.4em\relax New York, NY, USA: Association for Computing Machinery, Sep. 2020, pp. 189--190. [Online]. Available: \url{https://dl.acm.org/doi/10.1145/3387940.3392168}
\BIBentrySTDinterwordspacing

\bibitem{matthies_playing_2020}
\BIBentryALTinterwordspacing
C.~Matthies, ``Playing with your project data in scrum retrospectives,'' in \emph{Proceedings of the {ACM}/{IEEE} 42nd {International} {Conference} on {Software} {Engineering}: {Companion} {Proceedings}}, ser. {ICSE} '20.\hskip 1em plus 0.5em minus 0.4em\relax New York, NY, USA: Association for Computing Machinery, Oct. 2020, pp. 113--115. [Online]. Available: \url{https://dl.acm.org/doi/10.1145/3377812.3382164}
\BIBentrySTDinterwordspacing

\bibitem{kanban_online}
\BIBentryALTinterwordspacing
D.~Vacanti and J.~Coleman, ``{K}anban {G}uides.'' [Online]. Available: \url{https://kanbanguides.org/}
\BIBentrySTDinterwordspacing

\end{thebibliography}

\end{document}